\documentclass{ws-procs9x6}
\usepackage{latexsym}
\def\citebk#1{\mbox{[\hspace{0.9mm}\raisebox{-1.85mm}[0mm][0mm]
  {\Large\cite{#1}}\hspace{-0.1mm}]}}

\def\citebkcp#1{\mbox{[\hspace{0.8mm}\raisebox{-1.5mm}[0mm][0mm]
  {\large\cite{#1}}\hspace{-0.2mm}]}}

\newcommand{\be}{\begin{eqnarray}}
\newcommand{\ee}{\end{eqnarray}}
\newcommand{\beq}{\begin{equation}}
\newcommand{\eeq}{\end{equation}}
\newcommand{\Sigcon}{\langle \bar \psi \psi \rangle}
\newcommand{\mat}{\left ( \begin{array}{cc}}
\newcommand{\emat}{\end{array} \right )}
\newcommand{\matf}{\left ( \begin{array}{cccc}}
\newcommand{\nn}{\nonumber}
\newcommand{\cal}{\it} 
\begin{document}

\renewcommand{\thefootnote}{\fnsymbol{footnote}}
\title{Dirac Spectra and Real QCD at Nonzero Chemical Potential}
\author{D. Toublan$^1$ and J.J.M. Verbaarschot$^2$}
\address{$^1$Department of Physics, UIUC, \\1110 West Green Street,
Champaign-Urbana, Il 61801, USA\\ E-mail:toublan@uiuc.edu}
\address{$^2$Department of Physics and Astronomy,
SUNY at Stony Brook, \\
Stony Brook, NY 11794, USA\\
E-mail: verbaarschot@tonic.physics.sunysb.edu}

%\begin{normalsize}
%\begin{flushright}
%\begin{tabular}{l}
%SUNY--NTG--02/28\\
%July 2002
%\end{tabular}
%\end{flushright}
%\end{normalsize}

\maketitle

\abstracts{ 
We show that QCD Dirac spectra well below $\Lambda_{\rm QCD}$, both at
zero and at nonzero chemical potential, can be obtained from
a chiral Lagrangian. At nonzero chemical potential 
Goldstone bosons with nonzero baryon number condense beyond a
critical value. Such superfluid phase transition is likely to
occur in any system with a chemical potential with the quantum numbers
of the Goldstone bosons. We discuss the phase diagram for one such
system, QCD with two colors, and show the existence of 
a tricritical point in an effective potential approach.
}

\section{Introduction}
For strongly interacting quantum field theories such as QCD
a complete nonperturbative analysis from first principles
is only possible by means of large scale Monte Carlo simulations. 
Therefore, partial analytical results in some parameter domain
of the theory are extremely valuable, not only to provide additional
insight in the numerical calculations, but also as an independent
check of their reliability. This has been our main motivation 
for analyzing such domains. The principle idea we have been pursuing is based
on chiral perturbation
theory \citebk{Weinberg,GaL}: because of confinement and the
spontaneous breaking of
chiral symmetry,  the low-energy chiral
limit of QCD is a theory of weakly interacting Goldstone bosons
which are described by a chiral Lagrangian that is completely
determined by the symmetries of QCD. This idea can be applied
to the QCD Dirac spectrum which can be extracted from the valence
quark mass dependence of the chiral condensate. The
valence quark mass is not a physical parameter of the QCD partition
function
and can be chosen in a domain where
the valence quark mass dependence of the QCD partition function can
be described to an arbitrary accuracy by a corresponding chiral
Lagrangian. If the Compton wavelength of Goldstone bosons containing
only valence quark masses is much larger 
than the size of the box the low-energy effective
theory simplifies even much further \citebk{Vplb}. Then only the zero
momentum component of the Goldstone fields has to be taken into account so
that 
the valence quark mass dependence of the QCD partition function is
given by a
unitary matrix integral. 
This idea was first applied
to the QCD partition function \citebk{LS} with quark masses of order
$m \sim 1/V\Sigcon $ (with $\Sigcon$ the chiral condensate and $V$ 
the volume of space-time). However, we emphasize that for physical values
of the quark masses and volumes, a part of the Dirac
spectrum, as probed by the valence quark mass, is always in this
mesoscopic domain of QCD. More precisely, using the Gell-Mann-Oakes-Renner
relation (with $F$ the pion decay constant), in the domain
\be
\frac{m_v \Sigcon} {F^2} \ll 1/\sqrt V \quad {\rm and} \quad V^{1/4}\Lambda_{\rm
QCD} \gg 1,
\label{domain}
\ee
the kinetic term in the chiral Lagrangian can be ignored and the 
valence quark mass dependence of the 
QCD partition function reduces to a unitary matrix 
integral \citebk{Vplb,James}. This integral is equivalent to a chiral
Random Matrix Theory in the limit of large matrices \citebk{SV,V}.
The second condition ensures that
excitations of the order $\Lambda_{\rm QCD}$ decouple from the
low-energy sector of the partition function.

At nonzero baryon chemical potential the Dirac spectrum is scattered in 
the complex plane. However,
at a sufficiently small nonzero baryon chemical potential 
and finite physical quark masses, the Dirac spectrum in the phase of
broken chiral symmetry
is still described by a partition
function of Goldstone bosons containing valence 
quarks \citebk{TVspect}. In order to
eliminate the fermion determinant containing the valence quarks, one
has to calculate the valence quark mass dependence of the chiral 
condensate in the limit of a vanishing number of valence quarks.
The
existence of this limit requires the introduction of conjugate antiquarks
\citebk{Girko,Misha}, resulting in the appearance of Goldstone
bosons with nonzero baryon number containing only valence quarks.
They condense if the chemical potential exceeds
their mass. 
In terms of the Dirac spectrum this phase transition is visible as a
sharp boundary of the locus of the eigenvalues.

Such phase transition to a Bose condensed phase is likely to occur
in any
theory with a chemical potential with
the quantum numbers of Goldstone bosons.  For example, for QCD with
two fundamental colors \citebk{KST,KSTVZ}
or for adjoint QCD with two or more colors \citebk{KSTVZ}, the
lightest baryon is a Goldstone boson. 
A transition to a Bose
condensed phase occurs for a chemical potential larger than the mass of
this boson. Other examples are pion condensation, which may occur for a
nonzero isospin chemical potential \citebk{misha-son},
and kaon condensation which may occur
for a nonzero strangeness chemical potential \citebk{domstrange}. 
If the mass of the Goldstone bosons and the chemical potential are
both well below $\Lambda_{\rm QCD}$,
such phase transition can be described in terms of a chiral
Lagrangian. We have analyzed such Lagrangian for QCD with two colors at
nonzero temperature and chemical potential \citebk{KSTVZ,STV1,STV2}. In an
effective potential
approach we have found a tricritical point \citebk{STV2} in agreement with
recent 
lattice QCD simulations \citebk{domlat}.

We start this lecture by discussing QCD Dirac spectra at zero chemical
potential and explaining its description in terms of a chiral Lagrangian.
In section 3 we analyze QCD Dirac spectra at nonzero chemical potential.
The phase diagram of QCD with two colors at nonzero temperature and
chemical potential is discussed in section 4 and concluding remarks 
are made in section 5. 

\section{Dirac Spectrum at Zero Chemical Potential}

The Euclidean QCD Dirac operator is given by
\be
iD = \gamma_\mu (\partial_\mu + i A_\mu) ,
\ee
where the $\gamma_\mu$ are the Euclidean gamma matrices and the $A_\mu$ 
are $SU(N_c)$ valued gauge fields. The Dirac spectrum for a fixed
gauge field configuration is obtained by 
solving the eigenvalue equation
\be
iD \phi_k = i\lambda_k \phi_k.
\ee
In a regularization scheme with a finite number of $N$
eigenvalues, the average spectral density is defined by
\be
\rho(\lambda) = \langle \sum_{k=1}^N\delta(\lambda-
\lambda_k) \rangle_{\rm QCD} ,
\ee
where the average $\langle \cdots \rangle_{\rm QCD}$ 
is over gauge field configurations weighted by the
Euclidean QCD action. As a result of  the averaging we expect that
$\rho(\lambda)$ 
will be a smooth function of $\lambda$.
Because of the involutive automorphism $\gamma_5 iD \gamma_5 = -iD$ 
the Dirac operator can always be represented in block-form as
\be
iD = \mat 0 & iW \\ iW^\dagger & 0 \emat.
\ee
If $W$ is a square matrix the nonzero eigenvalues of $iD$ occur in pairs $\pm 
\lambda_k$. For nonzero topological charge the total number of zero eigenvalues
is given by the difference of the
the number of right-handed modes and left-handed modes. In that case,
the matrix $W$ is a rectangular matrix with the absolute value of the
difference between the
number of rows and columns equal to the topological charge. 
For very large values of $\lambda$  the Dirac spectrum converges to the
free Dirac spectrum so that the spectral density given by 
$\rho(\lambda) \sim V\lambda^3$. The smallest nonzero eigenvalue, 
$\lambda_{\rm min}$, is of the order of the average level spacing 
and is thus given
by
\be
\lambda_{\rm min} = \Delta \lambda = \frac 1{\rho(0)}.
\ee 

\subsection{Spontaneous Chiral Symmetry Breaking and Eigenvalue Correlations}

The chiral condensate is given by
\be
\langle \bar \psi \psi \rangle &=&
\lim_{\Lambda \to \infty} \lim_{m \to 0} \lim_{V\to \infty} 
\frac 1V \left\langle {\rm Tr} \frac 1{iD+m} \right\rangle_{\rm QCD} 
\nn\\
&=& 
\lim_{\Lambda \to \infty} \lim_{m \to 0} \lim_{V\to \infty} 
\frac 1V \int_0^\Lambda \frac {2m \rho(\lambda)}{\lambda^2 +m^2}.
\label{order}
\ee
The limit $m \to 0$ is taken before 
$\Lambda \to \infty$ to eliminate
divergent contributions from the ultraviolet part of the Dirac 
spectrum (the ultraviolet cutoff, $\Lambda$, may also appear in the
spectral density).
Because of spontaneous breaking of chiral symmetry, the limit $V \to
\infty$
cannot be interchanged with the limit $m \to 0$ in (\ref{order}).
If the chiral condensate is
nonzero the limits $ m \to 0^+$ and $m \to 0^-$ have opposite signs. This 
can only happen if $\rho(0) \sim V$. If we expand the spectral density as
\be
\rho (\lambda) = \rho(0^+) + a_1 |\lambda| + a_2 \lambda^2 + \cdots,
\ee
we obtain Banks-Casher formula \citebk{BC}
\be
\Sigcon = \lim_{V \to \infty}\frac{\pi \rho(0^+)}{V}.
\ee
In this article we avoid taking limits by mainly focusing  on  finite
values of $m$, $V$ and $\Lambda$.
  
Let us now consider the QCD partition function $Z(m_f)$,
\be
Z(m_f) = \langle \prod_f \prod_k(i\lambda_k +m_f) \rangle_{\rm YM},
\ee
where $\langle \cdots \rangle_{\rm YM} $ denotes averaging with respect
to the Yang-Mills action.
Because in the thermodynamic limit the derivative  of the partition function
with respect to $m_f$ has a discontinuity across the
imaginary axis, we expect that its zeros are also located on the
imaginary axis as well and, for finite volume,  are spaced as $1/V$. 
This average 
can also be written as an average over the joint eigenvalue
distribution 
\be 
\rho( \lambda_1, \lambda_2, \cdots )
\equiv \langle \delta(\lambda_1 - \lambda_1^A)\delta(\lambda_2 - \lambda_2^A)
\cdots \rangle_{\rm QCD},
\ee
where $\lambda_k^A$ are the eigenvalues of the Dirac operator for 
a given gauge field configuration $A$. This results in
\be
Z(m_f) = \int \rho(\lambda_1, \lambda_2, \cdots) \prod_f\prod_k (i\lambda_k 
+m).
\ee
If the eigenvalues are uncorrelated the joint eigenvalue distribution 
factorizes into one-particle distributions
\be
\rho(\lambda_1, \lambda_2, \cdots) = \rho_1(\lambda_1)\rho_1(\lambda_2) \cdots
\ee
and the partition function is the product of $N$ identical factors. For 
example, for $N_f = 1$, in the sector of zero topological charge, we obtain
 \be
Z(m) = (\langle \lambda^2 \rangle_1 + m^2)^N,
\ee
where $\langle \cdots \rangle_1$ is the average with respect to the
one particle distribution (which in this case is the average spectral
density of the QCD Dirac operator).
Therefore $Z(m)$ is a smooth function as $m$ crosses the imaginary axis
along the real axis and chiral symmetry is not broken. We conclude that
the absence of eigenvalue correlations implies that chiral symmetry is
not spontaneously broken, or conversely, 
if chiral symmetry is broken spontaneously the eigenvalues
of the Dirac operator are necessarily correlated. The question we wish
to answer is what are these correlations.

\subsection{Low Energy Limit of QCD}

Because of confinement 
the chiral limit of
QCD at low energy is a theory of weakly interacting Goldstone bosons. 
For small values of the quark masses $m_f$ and chemical potentials $\mu_f$ 
the QCD partition function coincides with a partition function of
Goldstone bosons:
\be
Z_{\rm QCD} (m_f, \mu_f, \theta) \sim Z_{\rm Gold}(m_f, \mu_f, \theta),
\ee
where $\theta$ is the vacuum $\theta$-angle.
Up to phenomenological coupling constants,
the mass dependence of $Z_{\rm Gold}$ is completely determined by the
symmetries and transformation properties of the QCD partition function. 
In particular, both partition functions have the same low mass
expansion. Equating the coefficients of powers of the quark masses leads
to sum-rules for the inverse Dirac eigenvalues \citebk{LS}.
To derive them we consider the Fourier components of the $\theta$
dependence which are just the partition function in a given sector of 
topological charge,
\be
Z_\nu(\cdots) = \frac 1{2\pi} \int_0^{2\pi} d\theta e^{i\nu\theta} 
Z(\cdots, \theta).
\ee
As an example, let us consider the case $N_f = 1$, $\mu_f =0$ and
$\nu = 0$. In this case there are no Goldstone bosons and the
the mass dependence of the partition function for $\theta = 0$ is given by
\be
Z \sim e^{V \Sigma (m + m^*)/2}.
\ee
The $\theta$ dependence id obtained from the substitution $m \to
me^{i\theta}$. For the sector of zero topological charge we thus find
\be
\langle(\lambda^2 + m^2) \rangle_{\nu = 0} &\sim& \frac 1{2\pi} \int_0^{2\pi}
d \theta e^{mV\Sigma \cos\theta} \nn \\
&=& 1 + \frac 14 m^2 V^2\Sigma^2 + \cdots.
\ee
This result in the sum rule \citebk{LS}
\be
\left\langle \sum_{\lambda_k > 0} \frac 1{\lambda_k^2}
\right\rangle_{\nu = 0}
= \frac{V^2 \Sigma^2}4.
\ee
In fact, an infinite number of sum rules can be derived for the 
partition function of QCD and QCD-like theories with spontaneous
chiral symmetry breaking
\citebk{LS,sell,SmV,lsmass,zyab}. Nevertheless, these sum rules are 
not sufficient to determine the Dirac spectrum. 

\subsection{Resolvent}
In order to derive the QCD Dirac spectrum we introduce the resolvent
\be
G(z) = \frac 1V \left\langle {\rm Tr} \frac 1{z+iD} \right\rangle_{\rm
QCD}.
\ee 
Here, $z$ is a complex 'valence quark mass' which does not occur inside
the fermion determinant that is included in the average.
The spectral density is obtained from the discontinuity of the resolvent
across the imaginary axis,
\be
\rho(\lambda) &=& \frac 1{2\pi} (G(i\lambda + \epsilon) - 
G(i\lambda - \epsilon))
\nn \\ &=& \frac 1{2\pi} (G(i\lambda + \epsilon) + G(-i\lambda + \epsilon)).
\ee
The resolvent can be obtained
\citebk{Brezin,Efetov,Morel,VWZ,pqChPT,Leung,OTV,DOTV}
from the generating function 
$Z_{\rm spect}(z,z',m_f)$,
\be
G(z) = \frac 1V \partial_z \left . Z_{\rm spect}(z, z',m_f)\right |_{z' = z},
\ee
with 
\be
Z_{\rm spect}(z,z',m_f) =
\left\langle \frac{\det(iD+z)}{\det(iD+z')}
\prod_f \det(iD+m_f) \right\rangle_{\rm YM}. 
\label{gen}
\ee
The variable $z$ is a parameter that probes the Dirac spectrum
and can  be chosen arbitrary small. For $ z,\, z',\, m_f 
\ll \Lambda_{\rm QCD}$ this 
partition function can be approximated arbitrarily well by a chiral 
Lagrangian which is completely determined by the symmetries of
the QCD partition function. In addition to fermionic quarks, this
partition function also contains bosonic ghost quarks. 
The corresponding
chiral Lagrangian therefore 
includes both bosonic and fermionic Goldstone bosons
with masses given by  $2 {\rm Re} (z) \Sigma/F^2$, 
${\rm Re} ( z+z') \Sigma/F^2$,
 ${\rm Re}( z +m_f)\Sigma/F^2$, etc., as 
given by the usual Gell-Mann-Oakes-Renner relation. 

The inverse fermion determinant can be written as a convergent
bosonic integral provided that ${\rm Re}(z) > 0$:
\be
\frac 1{\det(iD+z)} = \int d\phi d\phi^* e^{-\phi^*(iD + z) \phi}.
\ee
The convergence requirements restrict the possible symmetry
transformations
of the partition function. For example the axial $U(1)$ transformation
which in the fermionic case is given by
\be
\psi_R \to e^{i\theta} \psi_R, \qquad  \psi_L \to e^{-i\theta} \psi_L\nn \\
\bar \psi_R \to e^{-i\theta} \psi_R, \qquad  
\bar \psi_L \to e^{+i\theta} \psi_L,
\ee
would violate the complex conjugation structure of the bosonic integral
with $\bar\phi_R = \phi_L^*$ and $\bar \phi_L = \phi_R^*$. Instead, the
allowed $U_A(1)$ transformation is 
\be
\phi_R \to e^{s} \phi_R, \qquad  \phi_L \to e^{-s} \phi_L,\nn \\
\bar \phi_R \to e^{-s} \phi_R, \qquad  
\bar \phi_L \to e^{s} \phi_L,
\ee
with $s$ a real parameter. The Goldstone manifold is therefore {\it not} given
by the super-unitary group $SU(N_f+1|1)$ but rather by its complexified
version that reflects the convergence requirements of the bosonic
axial transformations \citebk{OTV,DOTV}. We will denote this manifold by 
$\hat{SU}(N_f+1|1)$ and
an explicit parameterization for the simplest case, $N_f =0$, will
be given below. 
Vector flavor symmetry transformations are consistent with the 
complex conjugation properties of the bosonic integral. This symmetry
group is thus given by $SU(N_f+1|1)$.

In the chiral limit the mass dependence of generating function (\ref{gen}) 
can be obtained from a chiral Lagrangian determined by its symmetries
and transformation properties. It is given by
\be
{\cal L} = {\rm Str} \partial_\mu U \partial_\mu U^{-1} -
\frac 12 \Sigcon {\rm Str}(M(U + U^{-1})), 
\ee
and the corresponding partition function reads
\be
Z_\nu = \int_{\hat{U}(N_f+1|1)} dU(x){\rm Sdet}^\nu U_0 e^{-\int d^4 x {\cal L}}.
\ee
Because $\nu$ is the global topological charge only the zero momentum 
component of $U$, denoted by $U_0$, appears in the argument of the
superdeterminant. In the chiral limit, QCD is flavor symmetric so that
the kinetic term of the chiral Lagrangian should be flavor symmetric as
well. Therefore, the pion decay constant of the extended flavor symmetry
is the same as in QCD. The mass matrix is given by
$M = {\rm diag}(m, \cdots, m, z, z')$.

If $ z \ll m_c \equiv F^2/\Sigcon \sqrt V$ the fluctuations of the zero
momentum modes are much larger that the fluctuations of the nonzero momentum
modes, which then can be ignored in the calculation of the resolvent. 
More physically, this condition means that the Compton wavelength of Goldstone
bosons containing ghostquarks with mass $z$ or $z'$ is much larger than 
the size of the box. In condensed matter physics, the energy scale $m_c$
is known as the Thouless energy and has been related to the inverse 
diffusion time of an electron through a disordered sample \citebk{Altshuler}.

In the Dirac spectrum we therefore can distinguish three different
energy scales,  the smallest
eigenvalue $\lambda_{\rm min}$, the Thouless energy $m_c$ and the 
QCD scale $ \Lambda_{\rm 
QCD}$. On mass scales well below $\Lambda_{\rm QCD}$ the mass dependence
of the QCD partition function is given by the chiral Lagrangian. For 
mass scales well below the Thouless energy only the zero momentum modes
have to be taken into account. However, for masses {\it not} much larger than
$\lambda_{\rm min}$, a perturbative calculation breaks down and 
the group integrals have to be performed exactly.
An interesting possibility is if $\lambda_{\rm min}$ and $m_c$
coincide which may lead to critical statistics \citebk{antonio}.

In the zero momentum limit, it is straightforward to calculate the integrals
over the superunitary group. The simplest case is the quenched case
($N_f =0$) where $U$ can be parameterized as
\be
U = \mat e^{i\theta} & \alpha \\ \beta & e^s \emat,
\ee
with $\alpha$ and $\beta$ are Grassmann variables, $\theta \in [0, 2\pi]$
and $ s \in \langle -\infty,\infty\rangle$.
In terms of the rescaled variable $u = z V \Sigcon$, one obtains the resolvent
\be
\frac {G(u)}{\Sigcon} = u (K_a(u) I_a(u) + K_{a-1}(u) I_{a+1}(u)) + \frac{\nu}u,
\label{resval}
\ee
where $ a= N_f + |\nu|$. From the definitions of the modified Bessel functions
it is clear that the compact/noncompact parameterization of the superunitary
group is essential. The microscopic spectral density is obtained from
the discontinuity of the resolvent and is given by
\be
\rho_s(\zeta) = \frac {\rho(\zeta/V\Sigcon)}{V\Sigcon}= 
\frac \zeta2 (J_a^2(\zeta)- J_{a+1}(\zeta) J_{a-1}(\zeta))
+\nu\delta(\zeta),
\label{resrho}
\ee
where $\zeta = \lambda V \Sigcon$.

\begin{center}                                                         
\begin{figure}[!ht]                                                   
\centering\includegraphics[width=55mm,angle=270]           
%{figcd.ps}                                                                
{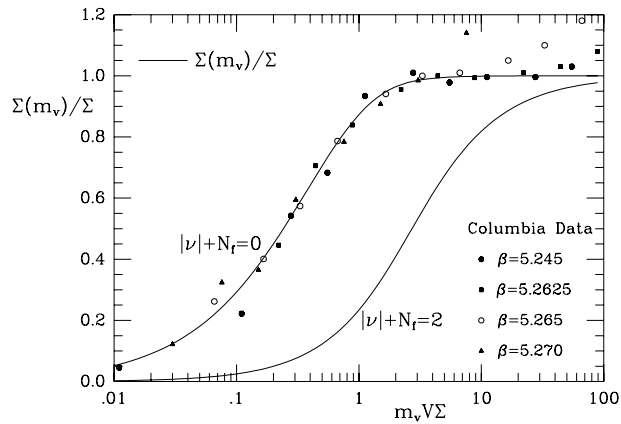}
\caption[]{The valence quark mass dependence of the chiral condensate   
$\Sigma(m_v)$ plotted as $\Sigma(z/m_v)/\Sigma$ versus $m_v V\Sigma$.  
The dots and                                                           
squares represent lattice results by the Columbia group \citebkcp{Chan95}    
 for values of $\beta$ as indicated in the label of the figure.
(Figure adapted from ref. \citebkcp{Vplb}).}     
\label{fig1}                                                          
\end{figure}                            
\end{center}

\subsection{Lattice Results}
The properties of the Dirac spectrum have been analyzed in many
lattice QCD simulations 
\citebk{Chan95,Vplb,l1,l2,l3,l4,l5,l6,l7,l8,l9,l10,l11,l12,l13,l14,l15} 
\citebk{l16,l17,l18,l19,l20,l21}
and have been found to be in complete agreement with the
conclusions of the previous section. We only show  three representative
examples.

In Fig. \ref{fig1} we show the valence mass dependence of the chiral
condensate as calculated by the Columbia group \citebk{Chan95}. In this
figure the valence quark mass is denoted by $m_v$ and
$\Sigma =\Sigcon$.  Our variable $u$
in
(\ref{resval}) is thus given by $m_v V \Sigma$ and $\Sigma(m_v)$
should be identified with $G(u)$. The reason that the lattice data
agree with the quenched approximation is that the sea-quark masses 
in the lattice calculation are much larger than the valence
masses. The topological charge is zero because the instanton zero
modes are completely mixed with the nonzero modes due to the 
lattice discretization. 
%{\it why is that so? D}
%This is so because of the so called Vink-Smit shift. Also numerically,
% if you just look at Dirac spectra, there is no sign of topology

%\begin{figure}[th]
%\epsfxsize=10cm   %width of figure - will enlarge/reduce the figures
%\epsfbox{fig3.eps}
%\figurebox{2cm}{3cm}{} %to have a box alone
%\centerline{\epsfxsize=3.5in\epsfbox{valkyoto.ps,angle=270}}
%\caption{The valence quark mass dependence of the chiral
%condensate for QCD with $N_f=2$ and $N_c=3$. 
%The analytical result (\ref{valres}) is
%represented by the  solid curves. 
%\label{fig1}}
%\end{figure}

\begin{figure}[!th]
%\epsfxsize=10cm   %width of figure - will enlarge/reduce the figures
%\epsfbox{fig3.eps}
%\figurebox{2cm}{3cm}{} %to have a box alone
\vspace*{3cm}
\centerline{\epsfxsize=8cm \epsfbox{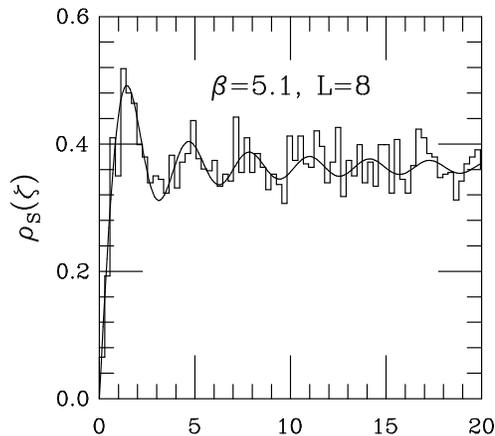}}
\vspace*{-5cm}
\caption[]{The microscopic spectral density for quenched QCD with three
colors. The solid curve represents the analytical result
(\ref{resrho}) for $a=0$.(Figure adapted from ref. \citebkcp{l9}) 
\label{fig2}}
\end{figure}

%\vspace{-2cm}
Because the valence quark mass dependence agrees with (\ref{resval})
the corresponding lattice QCD microscopic spectral density should
agree with (\ref{resrho}). This was shown by two independent
calculations \citebk{l9,l10}. In fig. \ref{fig2} we show
results for an $8^4$ lattice with quenched
staggered fermions \citebk{l9}.

\begin{figure}[!th]
%\epsfxsize=10cm   %width of figure - will enlarge/reduce the figures
%\epsfbox{fig3.eps}
%\figurebox{2cm}{3cm}{} %to have a box alone
\centerline{\epsfxsize=8cm\epsfbox{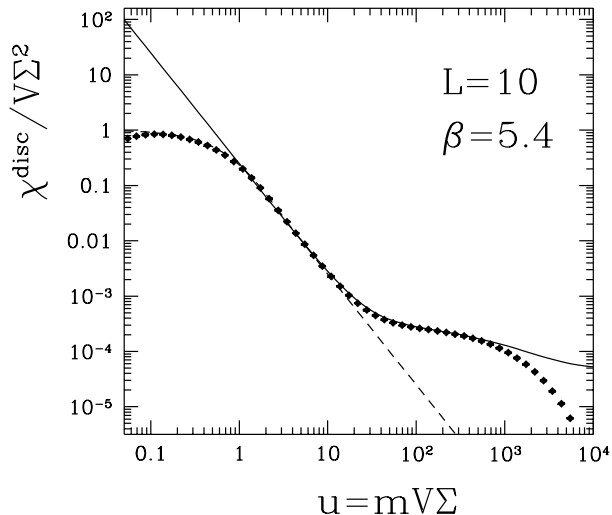}}
\caption[]{
The disconnected susceptibility for quenched SU(3) with staggered
      fermions (solid points).  The solid curve represents the
      prediction from  chPT, and the dashed one is the exact result for
      the zero momentum approximation to the chiral susceptibility.
      (Note the dashed line is hidden by the data points for $u<10$.)
      (Figure taken from ref. \citebkcp{l21}.)
\label{fig3}}
\end{figure}

%\vspace*{-1cm}
In Fig. \ref{fig3} we show the disconnected chiral susceptibility
defined by
\begin{equation}
  \chi^{\mathrm {disc}}(m_v)
    =\frac{1}{N}\left\langle\sum_{k,l=1}^N \frac{1}
      {(i\lambda_k+m_v)(i\lambda_l+m_v)}\right\rangle
     -\frac{1}{N} \left\langle\sum_{k=1}^N\frac{1}
      {i\lambda_k+m_v}\right\rangle^2 \:.
\end{equation}
This quantity can be obtained from the two-point spectral correlation
function 
but can also be directly computed in
chPT \citebk{VZ,domtwo,OTV,l21}. The dashed curve represents the 
result obtained from taking into account only the zero momentum modes
whereas the solid curve is obtained from a perturbative one-loop
calculation. Also in this figure $\Sigma = \Sigcon$.
This figure clearly demonstrates the existence of a
domain where a perturbative calculation can be applied to the zero
momentum sector of the theory.

\subsection{Chiral Random Matrix Theory}

Correlations of Dirac eigenvalues on the scale of the average level spacing
are completely determined by the zero mode part of the partition function
which only includes the mass term and the topological term of the
chiral Lagrangian. This raises the question of what is the most
symmetric theory that can be reduced to this partition function. The answer
is chiral Random Matrix Theory in the limit of large matrices. This
theory is invariant under and additional $U_R(n) \times U_L(n+\nu)$
group (with $n\times(n+\nu)$ the size of the nonzero blocks of the
Dirac matrix).
Because of this much larger symmetry group, all correlation function 
of the eigenvalues can be obtained analytically, often in a much 
simpler way than by means of the supersymmetric generating functions for the
resolvent. 

Before defining chiral Random Matrix Theories, we have to introduce the
Dyson index of the Dirac operator. It is defined as the number of independent
degrees of freedom per matrix element and is 
is determined by the anti-unitary
symmetries of the Dirac operator. They are of the form
\be
[AK,iD] =0,
\ee
with $A$ unitary and $K$ the complex conjugation operator. As shown by
Dyson \citebk{Dyso62}, 
there are only three different possibilities within an irreducible
subspace of the unitary symmetries
\be
i) &&{\rm There\,\, are\,\, no\,\, anti-unitary\,\, symmetries},\nn\\
ii) &&(AK)^2 = 1,\nn\\
iii) &&(AK)^2 = -1.
\ee
In the first case the Dirac operator is complex and the Dyson index is
$\beta_D = 2$. In the second case it is always possible to find a basis
in which the Dirac matrix is real and the Dyson index is $\beta_D =1$.
In the third case it is possible to express the 
matrix elements of the Dirac operator into 
selfdual quaternions and the Dyson index is $\beta_D = 4$.
The first case applies to QCD with three or more colors in the
fundamental representation. The second case is realized for QCD with
two colors in the fundamental representation, and the third case applies
to QCD with two or more colors in the adjoint representation. 

Chiral Random Matrix Theory is a Random Matrix Theory with the global
symmetries of the QCD partition function. It is defined by the partition
function \citebk{SV,V}
\be
Z_\nu(m_1, \cdots, m_f)= \int dW \prod_{f=1}^{N_f} {\det}^\nu(iD + m_f)
e^{- {\rm tr} V( W W^\dagger)},
\ee
where the Random Matrix Theory Dirac operator is defined by
\be
iD = \mat 0 & iW \\ i W^\dagger & 0 \emat,
\ee
and $ W$ is an $n\times (n+\nu)$ matrix so that $iD$ has exactly $\nu$ 
zero eigenvalues. In general the probability potential is a finite 
order polynomial. However, one can 
show \citebk{Brez96,senert,Akem97,Jack96,Guhr97,Kanz97} 
that correlations on the scale
of the average level spacing do not depend on the details of this
polynomial and the same results can be  obtained much
simpler from the Gaussian case. Depending on the Dyson index of the
Dirac operator we
have three different possibilities, the matrix elements of $W$ are
real, complex or self-dual quaternion for $\beta_D= 1, \,2,\, 4$, respectively.
The corresponding Gaussian ensembles are known as
the chiral Gaussian Orthogonal Ensemble (chGOE), the chiral Gaussian Unitary
Ensemble (chGUE) and the chiral Gaussian Symplectic Ensemble (chGSE),
in this order.
Together with the Wigner-Dyson Ensembles and four ensembles that can be
applied to superconducting systems, these ensembles can be classified according
to the Cartan classification of large symmetric spaces \citebk{class}. 

The results (\ref{resval}) and (\ref{resrho}) quoted in the previous
section were obtained first by means of standard Random
Matrix Theory methods \citebk{VZ,Vplb}.

\section{Dirac Spectra at Nonzero Chemical Potential}

Quenched lattice QCD Dirac spectra at $\mu\neq0$ were first obtained
numerically in the 
pioneering paper by Barbour et al. \citebk{everybody}
and have since then been studied in several other 
works \citebk{baillie,Tilomu,maria,simon}. 
Since the 
Dirac operator has no hermiticity properties at $\mu \ne 0$ its spectrum
is scattered in the complex plane. However, it was found \citebk{everybody}
that for not too large values of the chemical potential 
the spectrum is distributed homogeneously inside an oval shape with a
width  proportional to $\mu^2$. In this section we will explain these results
in terms of a chiral Lagrangian for phase quenched QCD at nonzero chemical
potential.

\subsection{Spectra of Nonhermitian Operators}
The spectral density of a nonhermitian operator is defined by
\be
\rho(\lambda) &=& \left\langle \sum_k \delta({\rm Re}(\lambda -
\lambda_k))
\delta({\rm Im}(\lambda - \lambda_k))\right\rangle_{\rm QCD} \nn \\
&=& \frac 1\pi \partial_{z^*} G(z),
\ee
where the resolvent $G(z)$ is defined by
\be
G(z) = \frac 1V \left\langle \sum_k \frac 1{z-\lambda_k} \right
\rangle_{\rm QCD}.
\ee
Often it is useful to interpret the real and imaginary parts of
the resolvent as the electric field in the plane at point  $z$ from
charges located at $\lambda_k$.

Since the fermion determinant is invariant for multiplication of the
Dirac operator by an unimodular matrix, one could analyze the spectrum
of various Dirac operators. The Dirac operator that is of interest is
the one with eigenvalues that are related to an observable. For example,
the Dirac operator in a chiral representation has the structure
\be
iD = \mat 0 & iW + \mu \\ iW^\dagger +\mu & 0 \emat.
\ee
In terms of its eigenvalues, the chiral condensate is given by
\be
\langle \bar \psi \psi \rangle = \langle \frac 1V 
\sum_k \frac 1{m +i\lambda_k}\rangle.
\ee 
If we are interested in the baryon number, on the other hand, we consider
the Dirac operator
\be
iD_\mu = \mat iW & m \\ m & i W^\dagger \emat,
\ee
which satisfies the relation $\det(iD_\mu + \mu) = \det(iD +m)$.
In terms of its eigenvalues $\mu_k$ the baryon density is given by
\be
n_B = \left\langle \frac 1V \sum_k \frac 1{\mu +i\mu_k} \right
\rangle_{\rm QCD}.
\ee
Finally, let us consider QCD at nonzero isospin chemical potential.
In this case the fermion determinant is given by
\be
\det \mat m &iW + \mu_I \\ iW^\dagger + \mu_I & m \emat
\det \mat m &iW - \mu_I \\ iW^\dagger - \mu_I & m \emat,
\ee
which can be rewritten as the determinant of the antihermitian matrix
\be
\matf 0 & 0&  -m& iW - \mu_I \\
   0 & 0 & iW^\dagger - \mu_I & -m \\
m & iW +\mu_I & 0 & 0 \\
iW^\dagger +\mu_I & m & 0 & 0 \emat.
\ee
In terms of its eigenvalues $i\pi_k$ , the pion condensate is given by
\be
\langle \pi \rangle = \left\langle \frac 1V \sum_k \frac 1{j_\pi +
i\pi_k} 
\right \rangle_{\rm QCD},
\ee
where $j_\pi$ is the source term for the pion condensate.

\subsection{Low Energy Limit of Phase Quenched QCD}
The generating function for  the quenched Dirac spectrum is given by
the replica limit $(N_f \to 0)$ of
phase quenched QCD partition function \citebk{Misha}  defined by
\be
Z &=& \langle [ \det(iD + z + \mu \gamma_0) {\det}^*(iD+z +\mu \gamma_0)
]^{N_f} \rangle_{\rm QCD} \nn \\ &=&
\langle [ \det(iD + z + \mu \gamma_0) \det(iD+z^* -\mu\gamma_0)
]^{N_f} \rangle_{\rm QCD}.
\label{phaseqpart}
\ee
Since this is a partition function of quarks and conjugate anti-quarks
we can have Goldstone bosons with nonzero baryon number. For a
chemical potential equal to half the pion mass we thus expect
a phase transition to a Bose condensed phase.
For a quark mass much less than $\Lambda_{\rm QCD}$ this  phase transition
can be described completely in terms a chiral Lagrangian. 
In nonhermitian Random Matrix Theory, the technique to determine the 
spectral density by analyzing a corresponding Hermitian ensemble is
known as Hermitization \citebk{Efetovnh,Feinnh},

The chiral Lagrangian is again determined by the symmetries and the
transformation properties of the QCD partition function. These can be
made more explicit if we rewrite the fermion determinant as
\be
\det \mat M_1 & d+B_R \\ -d^\dagger + B_L & M_2 \emat,
\ee
where $M_1 = M_2 = {\rm diag}(z, \cdots, z,z^*,\cdots, z^*)$ and
$B_L= B_R = {\rm diag}( \mu, \cdots, \mu, -\mu, \cdots, -\mu) $.
For $ z = \mu = 0$ our theory is invariant under $SU_L(2N_f) \times SU_R(2N_f)$.
For $z\ne 0$ and $\mu \ne 0$ this invariance can be restored if the 
the mass and chemical potential matrices are transformed as
\citebk{GaL,TVspect,misha-son}
\be
M_1 \to V_R M_1 V_L^{-1}, \qquad B_R \to V_R B_R V_R^{-1}, \\
M_2 \to V_L M_1 V_R^{-1}, \qquad B_L \to V_L B_L V_L^{-1} .
\ee
However, since $B_{R(L)}$ are a vector fields we can achieve local 
covariance by transforming them according to
\be
B_L \to V_L(\partial_0 + B_L) V_L^{-1},\nn \\
B_R \to V_R(\partial_0 + B_R) V_R^{-1}.
\ee
In the effective Lagrangian local covariance is obtained 
by replacing the derivatives in the kinetic term
 by a covariant derivative given by
\citebk{GaL}
\be
\partial_\nu \Sigma\to \nabla_\nu\Sigma \equiv \partial_\nu \Sigma
-B_L \Sigma + \Sigma B_R.
\ee
This results in the chiral Lagrangian
\be
\label{L2}
{\cal L} = \frac {F^2}4 {\rm Tr} 
\nabla_\nu \Sigma \nabla_\nu \Sigma^\dagger
- \frac G2 {\rm Tr}(M_1 \Sigma^\dagger+ M_2 \Sigma).
\ee
In our mean field analysis to be discussed below we only need the static
part of this Lagrangian which is given by \citebk{TVspect}
\be
{\cal L}^{stat} = \frac {F^2}4 \mu^2{\rm Tr} 
B_R\Sigma B_L\Sigma^\dagger
- \frac G2 {\rm Tr}(M_1 \Sigma^\dagger+ M_2 \Sigma).
\label{static}
\ee

\subsection{Mean Field Analysis}

In this subsection we describe the mean field analysis \citebk{TVspect} 
of the static Lagrangian (\ref{static}).
In phase quenched QCD, baryonic Goldstone modes contain a quark with
mass $z$ and a conjugate antiquark with mass $z^*$. According to the
GOR relation their mass is given by
\be
M^2 = \frac {(z+z^*)G}{2F^2}.
\ee
If the chemical potential is less than $M/2$ only the vacuum state
contributes to the QCD partition function. This results in
\be
Z = e^{V(z+z^*)G}.
\ee
We then find the following result for the resolvent and the spectral density
\be
G(z) = G, \qquad \rho(z) = 0, \qquad {\rm for }\qquad \mu < M/2.
\ee

\begin{center}                                                         
\begin{figure}[!ht]                                                   
\centering
\includegraphics[width=55mm,angle=270]{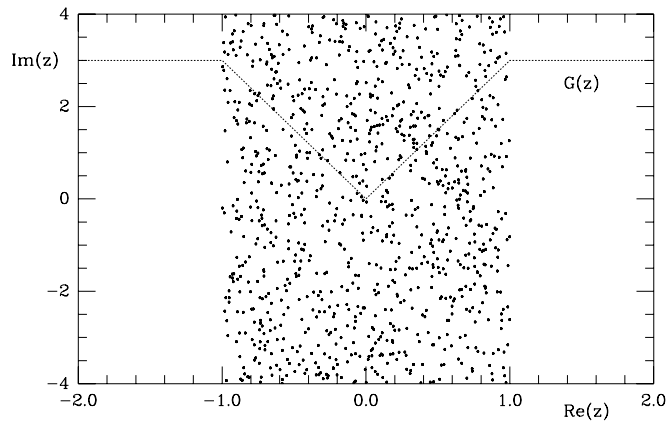}
 \caption[]{The distribution of eigenvalues of the Dirac operator in
the complex $z$-plane. The resolvent given by eq. (\ref{resmu}) is
represented by the dotted curve.
\label{fig6}}                                                          
\end{figure}                            
\end{center}

For $\mu > M/2$ the baryonic Goldstone modes condense resulting a 
nontrivial vacuum field which can be obtained from a mean field
analysis. The mass term and the chemical potential term in the static
Lagrangian are respectively  minimized by\footnote{The minimum
  $\Sigma_d$ is not
  unique which leads to massless 
Goldstone bosons in the condensed phase.}
\be
\Sigma_c = \mat 1 & 0 \\ 0 & 1 \emat, \; \; {\rm and} \; \; \Sigma_d =
\mat 0 & 1 \\ 1 & 0 \emat.
\ee 
%{\it Didn't  we prove that it was the true min in KSTVZ? D}
%But we did not prove it in this case. I vaguely remember that Misha and
% Son could not prove it in the isospin case.
A natural
ansatz for the minimum of the static 
Lagrangian (\ref{static}) is thus given by 
\be
\label{rot}
\Sigma = \Sigma_c \cos \alpha + \Sigma_d \sin \alpha.
\ee
An effective potential for $\alpha$ is obtained by substituting this
ansatz into the static Lagrangian. It is given by
\be
{\cal L}(\alpha) = \mu^2 F^2 N_f (\cos^2\alpha - \sin^2 \alpha) 
- G N_f (z+z^*) 
\cos \alpha.
\ee
This potential is minimized for $\bar\alpha$ given by 
\be
\begin{array}{lll}
\mu^2 < \frac {G(z+z^*)}{4 F^2}: &\quad \sin \bar\alpha = 0, 
\quad &{\cal L(\bar\alpha)} = -G N_f 
(z+z^*),\\
\mu^2 > \frac {G(z+z^*)}{4 F^2}: &\quad \cos \bar\alpha = \frac 
{G(z+z^*)}{4\mu^2F^2}, \quad &{\cal L(\bar\alpha)} = -\frac{G^2
N_f(z+z^*)^2}{8F^2\mu^2}.
\end{array}
\ee
~From the free energy at the minimum we easily derive the resolvent and
the spectral density (see Fig. \ref{fig6} in units with $2\mu^2 F^2/G
=1$)
\be
\begin{array}{lll}
\mu^2 < \frac {G(z+z^*)}{4 F^2}: &\quad G(z) = G, \quad &
\rho(\lambda) = 0,\\ 
\mu^2 > \frac {G(z+z^*)}{4 F^2}: &\quad G(z) = \frac{G^2 (z+z^*)}{F^2},
\quad &\rho(\lambda) = \frac {G^2}{4F^2 \mu^2}.
\end{array}
\label{resmu}
\ee
We conclude that the Dirac eigenvalues are distributed homogeneously 
inside a strip with width $\sim \mu^2$ in agreement with the numerical 
simulations \citebk{everybody}. For a discussion of correlations
of eigenvalues of a nonhermitian operator we refer to
the specialized literature
\citebk{Ginibre,fyodorov,Efetovnh,efetov4,forrester,Akemann,us}.  

In \citebk{misha-son} this analysis was applied to the problem of QCD at
finite isospin density with a partition function that coincides with
the phase quenched QCD partition function (\ref{phaseqpart})
\citebk{Wilczek}. 
In that
reference \citebk{misha-son}
it was also shown that the ansatz (\ref{rot}) is a
true minimum of the static Lagrangian. However, although we believe
that it is an absolute minimum, this has not yet been shown. 
%{\it cf  my remark above. D}

\section{Real QCD and Nonzero Chemical Potential}
The  analysis of the previous section can be repeated for any theory with 
a chemical potential with the quantum numbers of the Goldstone bosons.
Both for QCD with two colors in the fundamental representation and 
for QCD with two or more colors  in the adjoint representation,
a baryon has quark number two and is a boson. For broken chiral symmetry
some of these baryonic states are Goldstone bosons so that Bose-Einstein
condensation is likely to occur if the baryon
 chemical potential surpasses the mass of the Goldstone bosons.
For QCD with three or more colors in the fundamental representation we
expect a similar low energy behavior if we introduce a chemical potential
for isospin \citebk{Wilczek,misha-son} or strangeness \citebk{domstrange}
leading to pion condensation or kaon
condensation, respectively. Below we only discuss QCD with two colors.
 
\subsection{QCD with $N_c=2$}

\begin{figure}[!th]
\epsfxsize=10cm   %width of figure - will enlarge/reduce the figures
%\epsfbox{fig3.eps}
%\figurebox{2cm}{3cm}{} %to have a box alone
\centerline{\epsfxsize=6.0cm\epsfbox{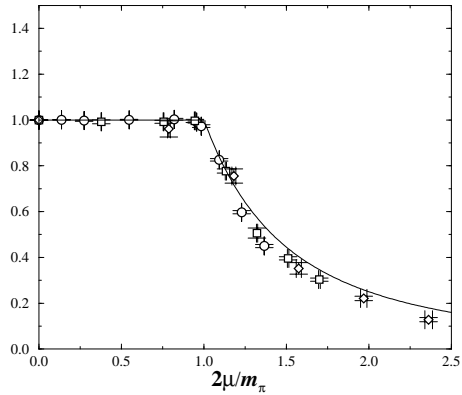}}
\caption[]{The chiral condensate versus $\mu/(m_\pi/2)$ for QCD with
two colors in the adjoint representation (Figure taken from
ref. \citebkcp{simonsuper}).
\label{fig5}}
\end{figure}

For simplicity, let us consider QCD with both two colors and two flavors.
In that case diquark mesons appear as flavor singlet. We thus have
five Goldstone bosons, three pions, a diquark and an 
anti-diquark\footnote{For QCD in the adjoint representation, 
the diquarks appear
as flavor triplet. For two flavors this results in three pions, three
diquarks and three anti-diquarks in agreement with spontaneous symmetry
breaking according to $SU(4) \to O(4)$}. Because $SU(2)$ is pseudo-real, the 
flavor symmetry group is enlarged to $SU(4)$. The quark-antiquark
condensate breaks this symmetry spontaneously to $Sp(4)$
\citebk{SmV,TV,symmLatt}.  
We can again write down a chiral Lagrangian based on this symmetry
group.
 Also in this case we find a competition between two condensates, and
in the Bose condensed phase, the chiral condensate rotates into
a diquark condensate for increasing values of the chemical potential
as in (\ref{rot}).
The mean field analysis proceeds in exactly the same way as in previous
section. For the chiral condensate we obtain
\be
\mu < m_\pi/2, \quad \langle \bar \psi \psi \rangle &=& 
\langle \bar \psi \psi \rangle_0, \nn \\ 
\mu > m_\pi/2, \quad \langle \bar \psi \psi \rangle &=& 
\langle \bar \psi \psi \rangle_0   \frac{m_\pi^2}{4\mu^2}.
\label{condmft}
\ee
In Fig. \ref{fig5}  we show that our predictions agree 
with lattice simulations
by Hands et al. \citebk{simonsuper}. 
The simulations were done for a $4^3 \times 8$ lattice
for $SU(2)$ in the adjoint representation and staggered fermions
which is in the same symmetry class as QCD with two colors in the
fundamental representation. A similar type of agreement was found by
several other groups \citebk{c1,c2,c3,c4,c5}

\begin{center}                                                         
\begin{figure}[!ht]                                                   
\centering
\includegraphics[width=45mm,angle=270]{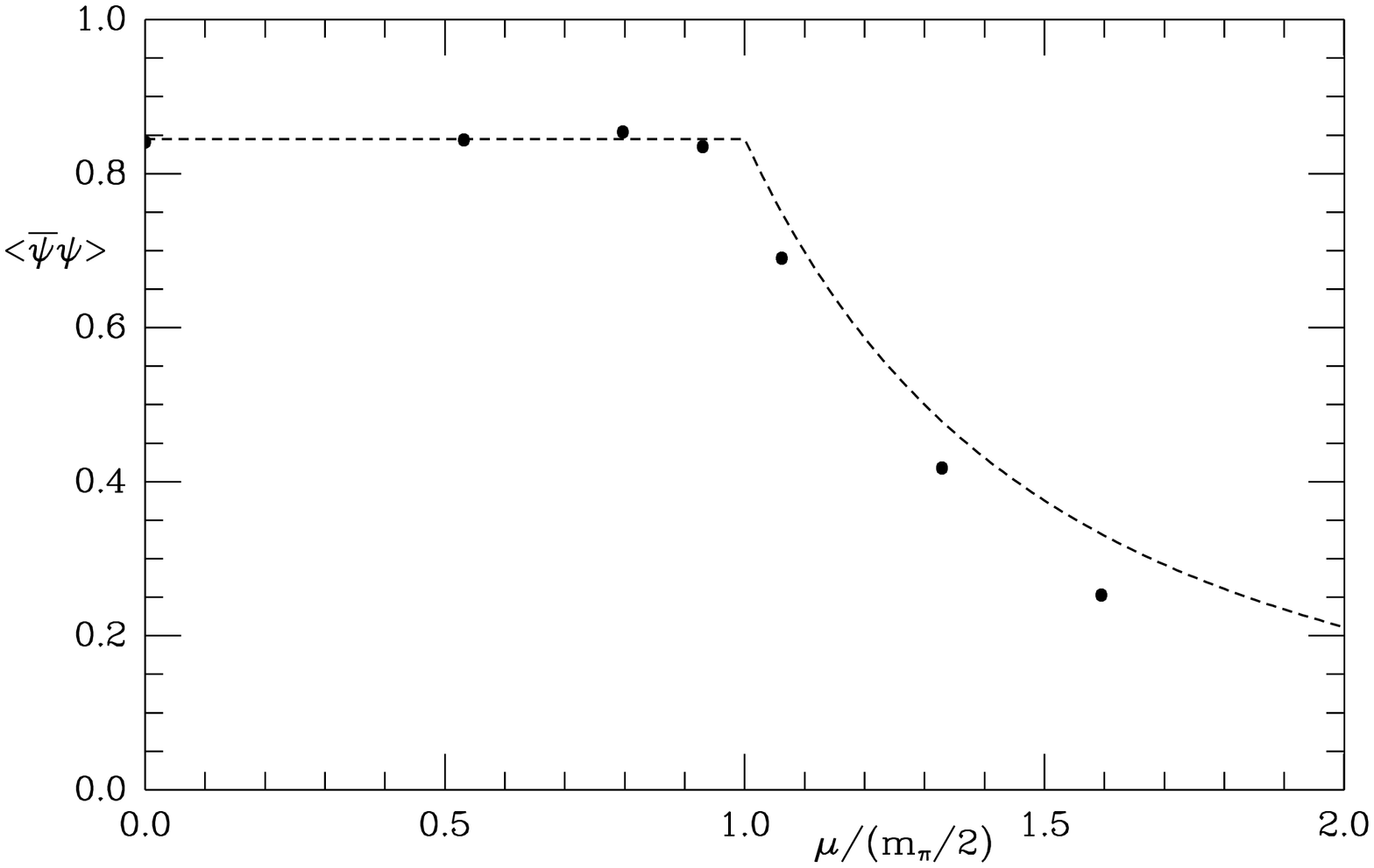}
\includegraphics[width=45mm,angle=270]{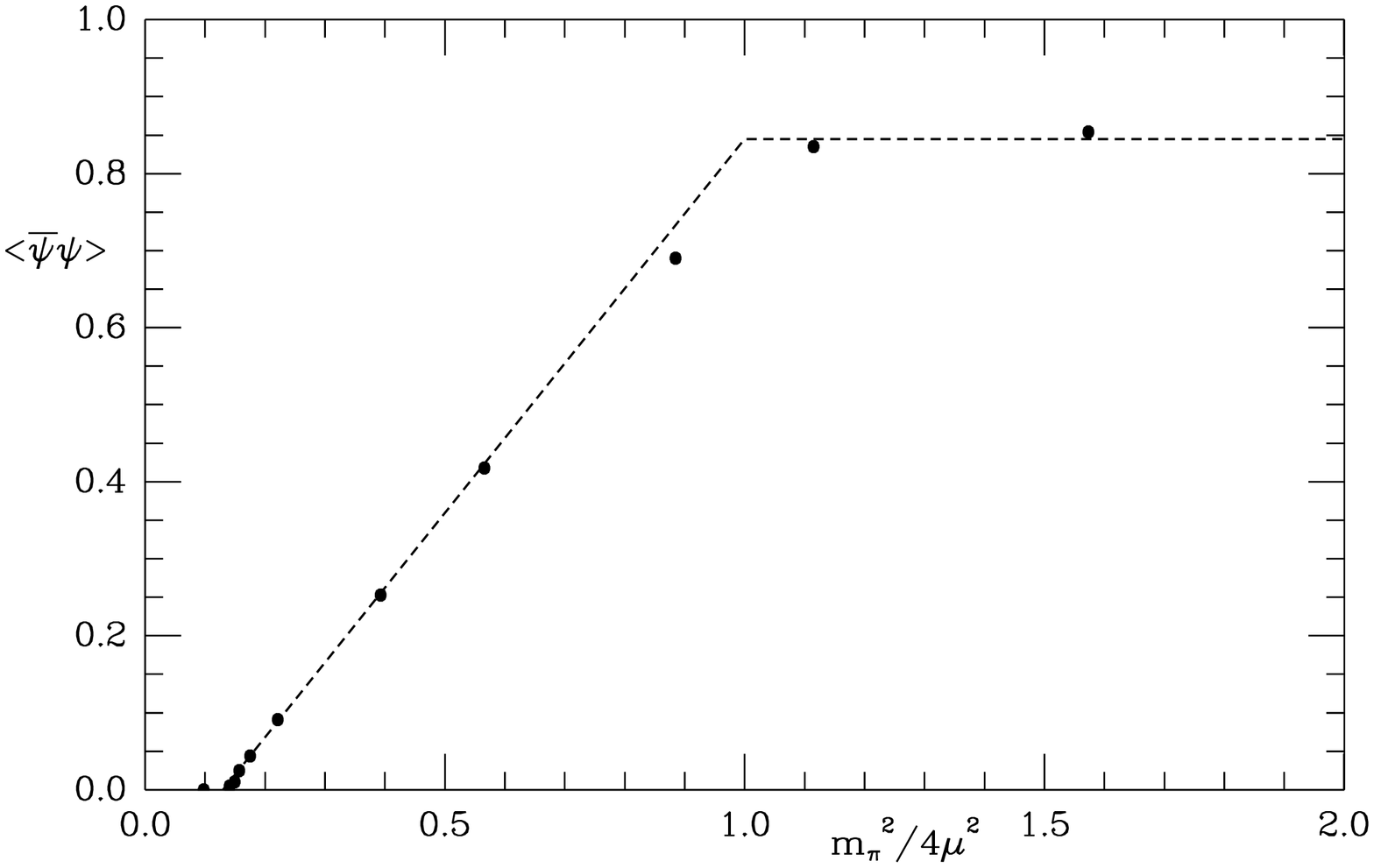}
\caption[]{The chiral condensate versus $\mu/( m_\pi/2)$ (upper) and
versus $m_\pi^2/4\mu^2$ (lower). The dashed curves in the lower figure
are drawn to guide the eye and in the lower figure they represent the
mean field result (\ref{condmft}) (Data points taken from ref.
\citebkcp{c3}).
\label{fig8}}                                                          
\end{figure}                            
\end{center}

Results for QCD with two colors in the fundamental representation 
obtained in \citebk{c3} are
shown in Fig. \ref{fig8}. Again we find good agreement with the mean
field results (\ref{condmft}). Furthermore, if we plot the same data 
versus $m_\pi^2/4\mu^2$ the curve reminds us of the resolvent for QCD
in phased quenched QCD after transforming the $z$ dependence of the
resolvent at fixed $\mu$ into a $\mu$ dependence at fixed $z$ (see
Fig. \ref{fig6}).  
Since the condensate can be interpreted as
the electric field at the quark mass
due to charges at the position of the eigenvalues
we have no eigenvalues for $\mu > m_\pi/2$ and for a narrow strip
along the $m_\pi^2/4\mu^2$  axis. 
In the remaining region the Dirac eigenvalues 
are distributed homogeneously. The absence of eigenvalues close
to the $m_\pi^2/4\mu^2$  axis is a signature\citebk{osbornmu} 
of $\beta_D =4$. 
Indeed, such behavior has been identified both 
numerically \citebk{osbornmu}
and analytically \citebk{efetov4}. 

%\begin{figure}[!th]
%\epsfxsize=10cm   %width of figure - will enlarge/reduce the figures
%\epsfbox{fig3.eps}
%\figurebox{2cm}{3cm}{} %to have a box alone
%\centerline{\epsfxsize=4.5cm\epsfbox{kogut1.ps}
%\epsfxsize=4.5cm\epsfangle=90\epsfbox{kogut2.ps}}
%\caption[]{First 3 normalized frequencies versus release location for
%clamped simply supported beam with internal slide
%release. \label{inter}}
%\end{figure}  

\subsection{Beyond Mean Field}

One of the recurring questions in the study of phase transitions is
the stability of the mean field analysis. In the following, we carry
out a
next-to-leading order study of the second order phase
transition found at the mean-field level. Additional details
can be found in \citebk{STV1,STV2}.
We will concentrate on the free
energy of the Bose condensed phase close to the mean-field critical
chemical potential $\mu_c=M/2$, with the leading order pion mass given
by the GOR relation: $M^2=G m_{\rm q}/F^2$.

The chiral Lagrangian ({\ref{L2}) contains the two operators that have the
lowest dimension in momentum space and 
that are invariant under local flavor transformations. There
are, of course, many operators of higher dimension that fulfill these
symmetry constraints. One has to introduce a systematic power counting
to account for their relative importance \citebk{GaL}.
Our power-counting scheme is the same as the one
used in chiral perturbation theory, extended to include the chemical
potential: $p \sim \mu \sim M \sim \sqrt{m_{\rm q}}$, where $p$ is
a Goldstone momentum. The leading order chiral Lagrangian contains
all the operators of order $p^2$ that fulfill the symmetry
constraints. The next-to-leading order chiral Lagrangian
contains all suitable operators of order $p^4$. In general, for
any $N_f$, there are ten such operators that contribute to the free
energy. They are made out of traces of $\nabla_\nu \Sigma \nabla_\nu
\Sigma^\dagger$ and of $M_1 \Sigma^\dagger +M_2 \Sigma$. The
next-to-leading order chiral Lagrangian can be written as
\begin{eqnarray}
  \label{L4}
  L^{(4)}=\sum_{i=0}^9 L_i {\cal O}_i.
\end{eqnarray}

At next-to-leading order, that is $p^4$, one has to take into account
the one-loop diagrams from the leading-order chiral Lagrangian
(\ref{L2}), as well as the tree diagrams from the next-to-leading
order Lagrangian (\ref{L4}). 
In this perturbative scheme, three Feynman diagrams contribute to the free
energy at next-to-leading order (see fig.~\ref{freeEnDiagr}). 
\begin{figure}[!ht]
\hspace{4cm}
\psfig{file=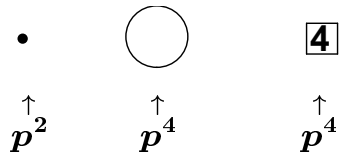,height=2cm,width=4.5cm}
\caption[]{Feynman diagrams that enter into the free energy at
  next-to-leading order. The dot denotes the contribution from $
    L$ (\ref{L2}), and the boxed $4$ the contribution from  $L^{(4)}$
    (\ref{L4}). The order in the momentum expansion is
    also given under each diagram.}
\label{freeEnDiagr}
\end{figure}
The one-loop diagram is divergent in four dimensions. The theory can be
renormalized by introducing renormalized coupling constants
\begin{eqnarray}
  \label{renorm}
  L_i  \rightarrow L_i^r(\Lambda) + \gamma_i \left[ \frac14
    -\Gamma(-d/2)  \right] \frac{\Lambda^{d-4}}{(4\pi)^{d/2}},
\end{eqnarray}
where $\Lambda$ is the renormalization scale and $\gamma_i$ are
numbers that can depend $N_f$ \citebk{GaL,STV1}. The renormalization
can be 
carried out order by order in the perturbation theory. It does not
depend on the chemical potential \citebk{STV1}.

The main technical difficulty at next-to-leading order comes from the
computation of the one-loop diagram in the Bose condensed
phase: Some modes are mixed \citebk{KSTVZ,STV1}. 
%The inverse 
%propagator of these mixed modes reads
%\begin{eqnarray}
%D^Q(p_0,\vec{p})=\left( \begin{array}{cc} p^2+M_1^2 & i M_3 \; p_0 \\
%    i M_3 \; p_0 & p^2+M_2^2 \end{array} \right), 
%\end{eqnarray} 
%where $M_1^2=M^2 \cos \alpha-4\mu^2 \cos 2\alpha$, $M_2^2= M^2 \cos
%  \alpha-4\mu^2 \cos^2\alpha $, and $M_3^2=16 \mu^2 \cos^2 \alpha$.
Because of this mixing, one-loop integrals 
%$\int d^dp
%\ln \det D^Q (p_0,\vec{p})$ 
may be quite complicated. 
However, we notice that the angle $\alpha$ that appears in (\ref{rot})
can be used as an order parameter of the Bose condensed phase. Since
we want to study the free energy of that phase near the
mean-field critical chemical potential $\mu_c=M/2$, it is sufficient to
compute the one-loop integrals for small $\alpha$, and $\mu$ close to
$M/2$. The free energy is then given by 
\begin{eqnarray}
  \label{freeEn}
  \frac{\Omega}{M^2 F^2} \sim {\rm cst} - \left( a_2+(2+a_3)
    (\frac{\mu}{M} -\frac12)  \right) \alpha^2+\left(\frac18-a_4 \right) \alpha^4
  + \dots
\end{eqnarray}
The coefficients $a_i$ come from the next-to-leading order
corrections. They are numbers that can be expressed in terms of
the renormalized coupling constants (\ref{renorm}). Their general form
is given by
\begin{eqnarray}
  a_i=\left(\sum_{k=0}^9 b_{ik} L^r_k(\Lambda) - \frac1{32 \pi^2}
    \sum_{k=0}^9 b_{ik} \gamma_k  \ln \frac{M^2}{\Lambda^2} \right)
  \frac{M^2}{F^2}. 
\end{eqnarray}
They do not depend on the renormalization scale $\Lambda$ and can 
be
evaluated from the $L^r_K$ which can in principle be obtained from lattice
simulations. 
They are expected to be 
small (of the
order of $0.05$ in 3-color QCD \citebk{GaL}).

The free energy (\ref{freeEn}) can be analyzed in the same way as a
Landau-Ginzburg model.  The coefficient of $\alpha^4$ is
positive. Therefore, there is a second order phase transition when the
coefficient of $\alpha^2$ vanishes. We thus find that the critical
chemical potential at next-to-leading order is given by
\begin{eqnarray}
  \label{critNLO}
  \mu_c=\frac12 M (1-a_2)=\frac12 m_\pi^{\rm NLO},
\end{eqnarray}
where $m_\pi^{\rm NLO}$ is the mass of the Goldstones at
next-to-leading order and at zero $\mu$. It is remarkable that the
next-to-leading order shift in $\mu_c$ corresponds exactly to the
next-to-leading order correction of $m_\pi$. At next-to-leading order,
we therefore find a second order phase transition at half the mass of
the lightest particle that carries a nonzero baryon charge. 

%It is rather straightforward to introduce a diquark source, $j$, in the
%chiral effective theory. This allows us to study the critical
%exponents at next-to-leading order in chiral perturbation theory. We
%find that
%\begin{eqnarray}
%  \alpha \sim (\mu-\mu_c)^{1/2} \;\; {\rm and \; \; at \; \;}
%  \mu=\mu_c: \; \alpha \sim j^{1/3}. 
%\end{eqnarray}
We have also calculated the critical exponents 
at next-to-leading order in chiral
perturbation theory and find that they are still given by their mean-field
values \citebk{STV1}. Since $d=4$ is the critical dimension beyond
which mean field 
exponents become valid, this is not entirely surprising.
From
the form of the propagators, we conjecture that the critical exponents
are given by mean-field theory at any (finite) order in perturbation
theory. 

In summary, we find that the next-to-leading order corrections are
only marginal. The main picture obtained from the mean-field analysis is
still valid at next-to-leading order: A second order phase
transition at $\mu_c=m_\pi/2$ with mean-field critical exponents
separates the normal phase from a Bose condensed phase.

\subsection{Nonzero Temperature}

At the one-loop level in chiral perturbation theory, the influence of
the temperature on the second-order phase transition can also
be studied \citebk{STV2,TempChPT}. In order to study the phase
transition at nonzero $T$ and 
$\mu$, we compute the free energy of the Bose condensed phase 
close the critical chemical potential $\mu_c=m_\pi/2$ at $T=0$. The
temperature dependence of the free energy is solely contained in the
1-loop diagram in fig.~\ref{freeEnDiagr}. 
%The main one-loop technical difficulties
%come from $\sum_n \int d^3p \ln \det
%D^Q(2\pi nT,\vec{p})$. 
Since we
are only interested in the behavior of the free energy of the Bose
condensed phase near the phase transition, it is sufficient to compute
it for small $\alpha$, small $T$, and $\mu$ close to $m_\pi/2$. This
procedure again leads to a free energy that can be analyzed as a usual
Landau-Ginzburg model. The minimum of the free energy is given by
\begin{eqnarray}
  \label{minT}
  \frac{\partial \Omega}{\partial \alpha}=0 \Rightarrow
\left\{
\begin{array}{l}
\alpha=0 \\
-c_2 +c_4 \alpha^2+c_6\alpha^4=0,
\end{array}
\right. 
\end{eqnarray}
where $c_i$ are coefficients that can be computed exactly. For
instance, we get that $c_2=( 32 \sqrt{2\pi^3} F_\pi^2 -
\zeta(\frac32) \sqrt{m_\pi T^3} ) ^2$ for $N_f=2$ \citebk{STV2}. We
find that at  
\begin{eqnarray}
  \label{tri}
  \mu_{\rm tri}&=&\frac{m_\pi}2 + \frac{m_\pi^3}{6\sqrt{3}
  \zeta^2(3/2) F_\pi^2} \left( 1-\frac{\zeta(1/2)
    \zeta(3/2)}{4 \pi}  \right)^{3/2}, \nonumber \\
T_{\rm tri}&=&2 m_\pi \frac{4\pi-\zeta(1/2) \zeta(3/2)}{3 \zeta^2(3/2)},
\end{eqnarray}
both $c_2$ and $c_4$ vanish. Therefore, we find that the second-order
phase transition line given by $c_2=0$, that is 
\begin{eqnarray}
  \label{2ndT}
  \mu_{\rm sec}(T)=\frac{m_\pi}2 + \frac{1}{32 F_\pi^2}
  \sqrt{ \frac{m_\pi^3 T^3}{2 \pi^3} } \zeta(3/2), 
\end{eqnarray}
ends at $\mu_{\rm tri}$. For a larger chemical potential, the phase
transition is of first order. The second-order phase transition line
(\ref{2ndT}) is in complete agreement with the semi-classical analysis
of a dilute Bose gas in the canonical ensemble. This phase diagram has
been confirmed by lattice simulations \citebk{c1,c4}.

\section{Conclusions}
Below $\Lambda_{\rm QCD}$ the QCD Dirac spectrum 
both at zero and at a sufficiently small 
nonzero chemical potential is described completely by
a suitable chiral Lagrangian. Below the Thouless energy, i.e. the scale for 
which the Compton wavelength of the Goldstone bosons is equal to the
size of the box, the Dirac spectrum can be obtained from the zero momentum
part of this theory. This matrix integral can also be derived from
a chiral Random Matrix Theory with the global symmetries of the QCD
partition function. Therefore, below the Thouless energy, the correlations
of QCD Dirac eigenvalues are given by chiral Random Matrix Theory.  
At nonzero chemical potential the Dirac eigenvalues are located inside
a strip in the complex plane. Going inside this strip the chiral
condensate rotates into a superfluid Bose-Einstein condensate. 
A very similar phase transition is found for any system with a chemical
potential with the quantum numbers of the Goldstone bosons. We have discussed
in detail the phase
diagram for QCD with two colors. Because of the Pauli-G\"ursey symmetry
diquarks appear as Goldstone bosons in this theory. We have analyzed
the phase diagram of this theory to one-loop order and have found   
a tricritical point in the chemical potential-temperature
plane. All our results are in agreement with recent lattice QCD
simulations.

\subsection{Acknowledgments}
Arkady Vainshtein is thanked for being a long lasting inspirational
force of our field and 
the TPI is thanked for its hospitality. 
D.T. is supported in part by
``Holderbank''-Stiftung and by NSF under grant NSF-PHY-0102409. 
This work was partially supported by the US DOE grant DE-FG-88ER40388.

\end{document}